\documentclass[10pt]{article}
\usepackage{amsmath}
\usepackage{amssymb}

\usepackage[dvips]{graphicx}

\def\aa{A\&A }

\def\aj{AJ }

\def\physletta{Phys. Lett. A}
\def\physrev{Phys. Rev. Lett.}
\def\soviet{Sov. Astron.}

\begin{document}

\setcounter{figure}{0}
\setcounter{table}{0}
\setcounter{footnote}{0}
\setcounter{equation}{0}

\vspace*{0.5cm}

\noindent {\Large PLAN FOR VLBI OBSERVATIONS OF CLOSE APPROACHES OF JUPITER TO COMPACT EXTRAGALACTIC RADIO SOURCES IN 2014-2016}
\vspace*{0.7cm}

\noindent\hspace*{1.5cm} A. GIRDIUK$^1$, O. TITOV$^2$\\
\noindent\hspace*{1.5cm} $^1$ Saint-Petersburg State University\\
\noindent\hspace*{1.5cm} Universitetsky prospect 28, Petrodvorets, 198504, Saint-Petersburg, Russia\\
\noindent\hspace*{1.5cm} e-mail: girduik@ipa.nw.ru\\
\noindent\hspace*{1.5cm} $^2$ Geoscience Australia\\
\noindent\hspace*{1.5cm} PO Box 378, Canberra, ACT 2601, Australia\\
\noindent\hspace*{1.5cm} e-mail: Oleg.Titov@ga.gov.au\\

\vspace*{0.5cm}

\noindent {\large ABSTRACT.} 
Very Long Baseline Interferometry is capable of measuring the gravitational 
delay caused by the Sun and planet gravitational fields. The post-Newtonian 
parameter $\gamma$ is now estimated with accuracy of $\sigma_{\gamma}=2\cdot 
10^{-4}$ using a global set of VLBI data from 1979 to present (Lambert, Gontier, 2009), 
and $\sigma_{\gamma}=2\cdot10^{-5}$ by the Cassini spacecraft (Bertotti et. al, 2003). 
Unfortunately, VLBI observations in S- and X-bands very close to the Solar limb (less than 2-3 degrees)
are not possible due to the strong turbulence in the Solar corona. Instead, the close approach of big 
planets to the line of sight of the reference quasars could be also used for testing of the 
general relativity theory with VLBI. Jupiter is the most appropriate among 
the big planets due to its large mass and relatively fast apparent motion 
across the celestial sphere. Six close approaches of Jupiter with quasars in 
2014-2016 were found using the DE405/LE405 ephemerides, including 
one occultation in 2016. We have formed tables of visibility for all 
six events for VLBI radio telescopes participating in regular IVS 
programs. Expected magnitudes of the relativistic effects to be measured 
during these events are discussed in this paper.

\vspace*{1cm}

\noindent {\large 1. USING VLBI OBSERVATIONS FOR TESTING GENERAL RELATIVITY}

\smallskip

Close approaches of the Sun and Jupiter to the apparent positions of compact extragalactic 
radio sources are used to estimate the PPN parameter $\gamma$ by the geodetic
VLBI technique. A first attempt to test the general relativity theory using the close 
pass of Jupiter to quasar 0201+113 has been done in 1988 
(Schuh et al., 1988) at the angular distance of 3$'$.5.
A more famous experiment was arranged on 8 Sep 2002 
(Fomalont \& Kopeikin, 2003) when Jupiter approached quasar 
J0842+1835 at the angular distance of 3$'$.7. Variations of the relative
separation on the sky between this quasar and a reference radio source were 
measured by the VLBA network and the Effelsberg 100-meter radio telescope.
Another experiment was done on 18 November, 2008 as a part of the 
session OHIG60 arranged by the International VLBI Service. 
During this session Jupiter approached quasar 1922-224 at an angular 
distance of 1$'$.2. Four VLBI stations observed this event for
about 12 hours.

\vspace*{0.7cm}

\noindent {\large 2. ESTIMATION OF THE PPN-PARAMETER $\gamma$ FROM THE VLBI OBSERVATIONS}

\smallskip

Besides three classical tests, the fourth test of general relativity - 
the delay of a signal propagating in the gravitational field, has been proposed by Shapiro (1964)
and known as the Shapiro delay. The difference between two Shapiro delays as
measured with two radio telescopes gives a gravitational delay which must
be considered at the standard reduction of the high-precision geodetic VLBI data.
The IERS Conventions 2010 (Petit and Luzum, 2010) comprises the 'consensus' 
formula for the gravitational delay which is valid for the most cases unless
a distant quasar and a deflecting body are too close. This formula is presented 
as follows

\begin{equation}\label{shapiro}
{\bigtriangleup t}_{grav}=\frac{(\gamma+1)GM}{c^3}\ln \frac{|\vec{r_1}|+\vec{s}\cdot\vec{r_1}}{|\vec{r_2}|+\vec{s}\cdot\vec{r_2}} ,
\end{equation}
where $\gamma$ - the PPN-parameter of general relativity (Will, 1993), $G$ - the 
gravitational constant, $M$ - the mass of gravitational body, 
$c$ - speed of light, $\vec{s}$ - the barycentric unit vector towards 
the radio source and $\vec{r_i}$ - 
the vector between gravitatiting body's center of mass and $i$-th telescope (see i.e. Kopeikin (1990),
Hellings and Shahid-Saless (1991), Klioner (1991) for more details).

An expression that links the gravitational delay and the formula for the light deflection angle (Einstain, 1916)
yet to be developed. To obtain it we have expanded the gravitational delay using the 
Taylor times series on $o(\frac{b}{r})$. Finally, the main terms 
of this expansion are given by

\begin{equation}\label{formula}
{\tau}_{grav}=-\frac{(\gamma+1)GM}{c^{3}}\frac{b}{r}\cos\varphi+\frac{(\gamma+1)GM}{c^{3}}\frac{b}{r}\frac{\;\sin\varphi\sin\theta\cos A}{1-\cos\theta}+o(\frac{b^2}{r^2}),
\end{equation}
where vectors $\vec{b}$, $\vec{r}$, and angles $\varphi$, $\psi$, $\theta$ and $A$ are shown on Fig. 1 (Turyshev, 2009).

\begin{figure}[h]
\begin{minipage}[h]{0.49\linewidth}
\begin{center}
\includegraphics[scale=0.25]{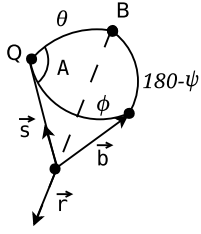}
\end{center}
\end{minipage}
\hfill
\begin{minipage}[h]{0.49\linewidth}
\caption{Schematic image shows positions of the quasar Q, deflecting body B (Jupiter), baseline vector $\vec{b}$, $\vec{r}$ - vector from Jupiter to geocenter, barycentric unit vector $\vec{s}$ to the quasar Q. Angle $\theta$ - the impact parameter, angle $\varphi$ between vectors $\vec{b}$ and $\vec{s}$, angle $\psi$ between vector $\vec{b}$ and $\vec{r}$}
\end{minipage}
\end{figure}

Surprisingly, we found the first term in (\ref{formula}) is equal to the term 
including the PPN parameter $\gamma$ at the geometric delay, but with opposite 
sign. Keeping in mind that ($\vec{b}\cdot\vec{s}=|\vec{b}|\cos\varphi$)
the formula for the total group delay recommended by the IAU (Eubanks, 1991), Petit and Luzum (2010):

\begin{equation}
\tau_{group}= \frac{\tau_{grav}-\frac{\vec{b}\cdot\vec{s}}{c}(1-\frac{(\gamma+1)GM}{c^2r}+...)}{1+\frac{1}{c}(\vec{s}\cdot(...))}=\frac{\tau_{\text{{\tiny GR}}}+...}{1+\frac{1}{c}(\vec{s}\cdot(...))},
\end{equation}
where $\tau_{\text{\tiny GR}}$ is the resultant contribution of the general relativity (GR) effects to 
the $\tau_{group}$,
includes two relavistic terms which cancel each other out. 
Then, $\tau_{\text{\tiny GR}}$ may be written as follows 

\begin{equation}\label{gr_1}
\tau_{\text{\tiny GR}}=\frac{(\gamma+1)GM}{c^3}\ln \frac{|\vec{r_1}|+\vec{s}\cdot\vec{r_1}}{|\vec{r_2}|+\vec{s}\cdot\vec{r_2}}  + \frac{(\gamma+1)GM(\vec{b}\cdot\vec{s})}{c^3r}
\end{equation}
or, from (2) and (4)

\begin{equation}\label{gr_2}
\tau_{\text{\tiny GR}}=\frac{(\gamma+1)GM}{c^{3}}\frac{b}{r}\frac{\;\sin\varphi\sin\theta\cos A}{1-\cos\theta}+o(\frac{b^2}{r^2}).
\end{equation}

Given that $\gamma =1$ in general relativity, and ingoring $o(\frac{b^2}{r^2})$ for the sake of simplicity

\begin{equation}\label{gr_calc}
\tau_{\text{{\tiny GR}}}=\frac{2GM}{c^3}\frac{b\;\sin\varphi\sin\theta\cos A}{r(1-\cos\theta)}.
\end{equation}

For the approximation of small angles (if $\theta\rightarrow 0$), (6) comes down to

\begin{equation}\label{gr_calc}
\tau_{\text{{\tiny GR}}}=\frac{4GM}{c^3}\frac{b\;\sin\varphi\cos A}{R},
\end{equation}
where $R = \theta \cdot r$ is the linear impact parameter.
It is now easily to note that the term (7) corresponds to the formula of the light deflection developed by Einstein in 1916:

\begin{equation}\label{alpha_clas}
\alpha''=\frac{4GM}{c^2R},
\end{equation}
\begin{figure}[h!]
\begin{minipage}[h]{0.49\linewidth}
\begin{center}
\includegraphics[scale=0.9]{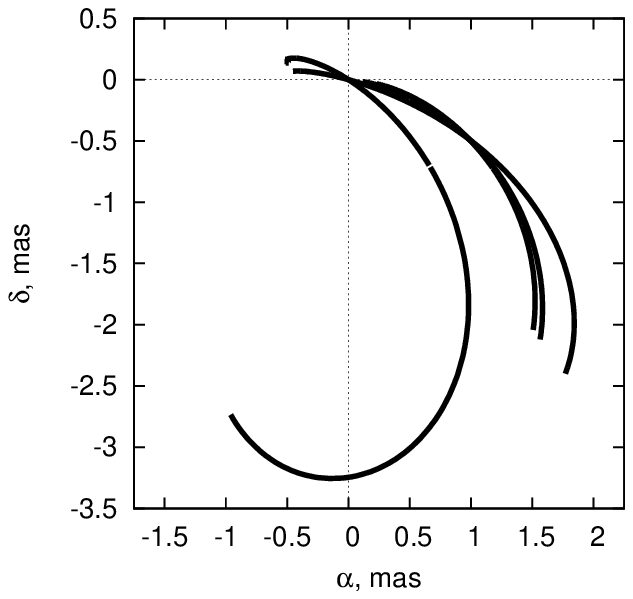}
\end{center}
\end{minipage}
\hfill
\begin{minipage}[h]{0.49\linewidth}
\begin{center}
\includegraphics[scale=0.9]{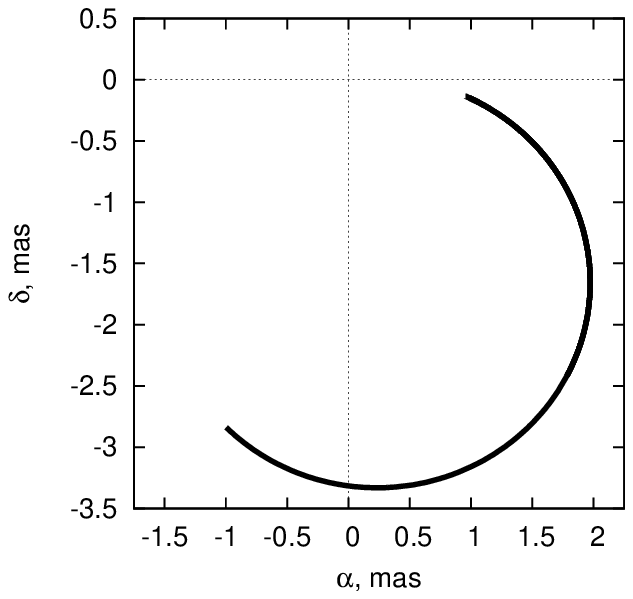}
\end{center}
\end{minipage}
\caption{$\alpha''_{A}$ for baselines: Hobart26-Tsukub32, Parkes-Tsukub32, Hobart26-Parkes, Hobart26-Kokee
on the left plot, and $\alpha''$ on the right plot.}
\end{figure}

as follows

\begin{equation}\label{al_csnA}
\tau_{\text{{\tiny GR}}}=\alpha''\frac{b}{c}\sin\varphi\cos A.
\end{equation}

Formula (\ref{al_csnA}) proves that in the first approximation the deflection angle (\ref{alpha_clas}) 
as measured with the geodetic VLBI is independent on a baseline length. The factor $\cos A$ 
is individual for each baseline Fig.2, therefore, implication of the angle $A$ in (\ref{al_csnA}) is important.

Fig. 2 displays the dependence of the factor $\cos A$ on the calculation of the deflection angle $\alpha''$ in 
the plane of equatorial coordinates ($\alpha,\delta$). The left plot shows the angle 

\begin{equation}
\alpha''_{A}=\alpha''\cdot\cos A, \;\;\; \alpha''_{A} = \frac{c\tau_\text{{\tiny GR}}}{b\sin\varphi}
\end{equation}
for four individual baselines (from the OHIG60 experiment on 18 November, 2008), where 
$\tau_{\text{{\tiny GR}}}$ is calculated by (\ref{gr_calc}). The right plot shows the angle as

\begin{equation}
\alpha'' = \frac{c\tau_\text{{\tiny GR}}}{b\sin\varphi\cos A}.
\end{equation}

All four individual curves shown on the 
left plot are merging to one common pathway on the right 
plot. Thus, it proves that the angle $A$ must be taken into account for proper conversion from 
$\tau_{\text{{\tiny GR}}}$ to $\alpha''$ and vice versa as (9) and (11). This is also true for an arbitrary
angle $\theta$ in (6).

\vspace*{0.7cm}

\noindent {\large 3.FUTURE EVENTS AND THEIR SCHEDULING}

\smallskip
\begin{table}[h!]
\begin{center}
\begin{tabular}{cccccc|c}
\hline
 N & quasar & date & $\theta''$ & Flux, mJy & $\frac{\alpha''}{\cos A}$, mas & small terms, pks, for b=6000 km\\
\hline
\hline
1 & 0846+184 & 28.08.2014 & 45 & ? & 6.0 & 8.5 \\
2 & 0918+167 & 09.02.2015 & 30 & ? & 12.8 & 39.1 \\
3 & 0912+171 & 24.05.2015 & 39 & ? & 7.7 & 14.2 \\
4 & 0920+168 & 08.06.2015 & 37 & ? & 7.8 & 14.8 \\
5 & 1109+070 & 26.03.2016 & 26 & $\approx 200$ & 14.5 & 50.1 \\
6 & 1101+077 & 09.04.2016 & 20(8) & $\approx 150$ & 16(46) & 142(503) \\
\hline
\end{tabular}
\caption{ Close approaches of Jupiter to quasars in 2014-2016.}
\end{center}
\begin{tabular}{l|p{0.02\linewidth}p{0.02\linewidth}p{0.02\linewidth}p{0.02\linewidth}p{0.02\linewidth}p{0.02\linewidth}p{0.015\linewidth}p{0.015\linewidth}p{0.02\linewidth}p{0.02\linewidth}p{0.02\linewidth}p{0.02\linewidth}p{0.02\linewidth}p{0.02\linewidth}p{0.02\linewidth}p{0.02\linewidth}}
\hline
angular distance$''$ & 170 & 148 & 126 & 104 & 83 & 64 & 48 & 41 & 46 & 61 & 79 & 100 & 121 & 143 & 166 \\
\hline
station/UT & 4 & 5 & 6 & 7 & 8 & 9 & 10 & 11 & 12 & 13 & 14 & 15 & 16 & 17 & 18 \\
\hline
\hline
Badary&+&+&+&+&+&+&+&+&+&+&+&+&+& &\\
Svetloe, Zelenchuk& & & &+&+&+&+&+&+&+&+&+&+&+&+\\
Medicina, Hartrao, Yebes& & & & & &+&+&+&+&+&+&+&+&+&+\\
Kokee, MN-VLBA, Seshan25&+&+&+&+&+&&&&&&&&&&\\
FD-LA-KP-OV-BR-VLBA, Pietown&+&+&&&&&&&&&&&&+&+\\
SC-HN-NL-VLBA&&&&&&&&&&&&+&+&+&+\\
Parkes, Tsukub32&+&+&+&+&+&+&+&+&&&&&&&\\
\hline
\end{tabular}
\caption{\label{tab:stan} Table of visibility for event on 08 June, 2015.}
\end{table}

Six close approaches of Jupiter to quasars will happen in 2014-2016 including one occultation.
Table 1 shows names of quasars, dates of event, impact parameters $\theta$, flux densities (X-band), 
the maximum angles of the light deflection for the major term and the contributions of the minor terms 
$o(\frac{b^2}{r^2})$ in (\ref{formula}) for baseline length of $b=6000$ km and $\gamma=1$. 
A total occultation of the Jupiter by the 
quasar 1101+077 will happen on 9 April, 2016. Therefore, the numbers for the minimum angle $\theta$  
and the angle $\theta$  at the limb of Jupiter are shown separately for that event. Flux densities of four 
quasars from Table 1 are not available presently and should be measured as soon as possible.

Table \ref{tab:stan} presents a visibility chart for several VLBI radio telescopes 
for the event on 8 June, 2015. This event will last than more ten hours (when the angular distance 
between objects is less then $3'$). The sign '+' notes the hour when both objects will be above the 
level of horizon for a particular VLBI station. Due to very weak flux for the quasars in Table 1, 
we have preselected only large size radio telescopes to ensure a reasonal integration time (300 seconds). 
Thus, many small VLBI radio telescopes were discarded from Table 1 after repeated calculations of the
integration time.

\vspace*{0.7cm}

\noindent {\large 4. PLAN FOR VLBI OBSERVATIONS}

\smallskip

The maximum deflection angle for the list of close approaches in Table 1 is 16 mas, or equivalent to
the time delay of about 1600 pks for a baseline of 6000 km length. When the current accuracy of a single group
delay as measured by VLBI is about 30 pks, the relative accuracy of $\gamma$ would reach 
$\sigma_{\gamma}  = 0.02$. Conservatively, it would be possible to evaluate the parameter $\gamma$ 
with accuracy $\sigma_{\gamma}  = 10^{-3}$ for a single VLBI experiment, if a substantial number of 
VLBI group delays is collected. In addition, it may be possible for one to prove existence of the minor terms 
from the last column of Table 1 for the two last approaches. It is necessary to organize as many large
radio telescopes as possible to maximize a potential amount of observations. Therefore, we are planning 
to submit proposals for the large available networks (IVS, VLBA, LBA) to observe these rare events.

\vspace*{0.7cm}

\noindent {\large 5. REFERENCES}

%
%
%
%
%
{

\leftskip=5mm
\parindent=-5mm

\smallskip

Schuh, H., Fellbaum, M., Campbell, J., Soffel, M., Ruder, H., Schneider, M., 1988, "On the deflection of radio signals in the gravitational field of Jupiter", \physletta, vol. 129, Issue 5-6, pp. 299-300.

Fomalont, E., Kopeikin, S., 2003, "The measurement of the light deflection from Jupiter: experimental results", \aj, vol. 598, Issue 1, pp. 704-711.

Einstein, A., 1916, "Die Grundlage der allgemeinen Relativitatstheorie", Annalen der Physik, vol. 354, Issue 7, pp. 769-822

Lambert, S., Gontier, A., 2009, "On radio source selection to define a stable celestial frame", \aa, vol. 493, Issue 1, 2009, pp. 317-323 

Bertotti, B., Iess, L., Tortora, P., 2003, "A test of general relativity using radio links with the Cassini spacecraft", Nature, vol. 425, Issue 6956, pp. 374-376

Shapiro, I., 1964, "Fourth Test of General Relativity", \physrev, vol. 13, Issue 26, pp. 789-791

Shapiro, S., Davis, J., Lebach, D., Gregory, J., 2004, "Measurement of the Solar Gravitational Deflection of Radio Waves using Geodetic Very-Long-Baseline Interferometry Data, 1979-1999", \physrev, vol. 92, Issue 12, id. 121101

Hellings, R., Shahid-Saless, B., 1991, "Relativistic Effects on VLBI Observables and Data Processing Algorithms", 
Proceedings of the U. S. Naval Observatory Workshop on Relativistic Models for Use in Space Geodesy, U. S. Naval Observatory, Washington, D. C., pp. 164-174.

Klioner, S., 1991, "General Relativistic Model of VLBI Observables", 
Proceedings of the U. S. Naval Observatory Workshop on Relativistic Models for Use in Space Geodesy, U. S. Naval Observatory, Washington, D. C., pp. 188-202.

Will, C., 1993, "Theory and Experiment in Gravitational Physics", Cambridge University Press

Eubanks, T. (ed.), 1991, Proc. of the U. S. Naval Observatory Workshop on Relativistic Models for Use in Space Geodesy, U. S. Naval Observatory, Washington, D. C.

Kopeikin, S., 1990, "Theory of relativity in observational radio astronomy", \soviet, vol.34, pp. 5-9

Turyshev, S., 2009, "Relativistic gravitational deflection of light and its impact on the modeling accuracy for the Space Interferometry Mission", Astronomy Letters, vol. 35, Issue 4, pp. 215-234

Petit, G., Luzum, B. (eds.), 2010, IERS Conventions (2010), IERS Technical Note 36, Frankfurt am Main: Verlag des Bundesamts fur Kartographie und Geodasie.


}

\end{document}